\documentclass[traditabstract]{aa} 
\usepackage{natbib}
\bibpunct{(}{)}{;}{a}{}{,}
\usepackage{graphicx}
\usepackage{revsymb}    
\usepackage{amsmath}    
\usepackage[usenames]{color}    
\usepackage{epstopdf}   
\def\ind#1{_{\rm#1}}

\newcommand{\vt}{\rm v}

\newcommand{\derivp} [2] {\frac {\partial #1 } {\partial #2} }
\newcommand{\deriv} [2] {\frac {\textrm{d} #1 } {\textrm{d} #2} }

\newcommand{\eq}[1] {Eq.\,(\ref{#1})}

\usepackage{txfonts}

\usepackage[switch]{lineno} 

\begin{document}

\title{Angular momentum redistribution by mixed modes \\ in evolved low-mass stars}
\subtitle{II. Spin-down of the core of red giants induced by mixed modes}

\author{K. Belkacem\inst{1}, J.P. Marques\inst{2}, M.J. Goupil\inst{1}, B. Mosser\inst{1}, T. Sonoi\inst{1}, R.~M. Ouazzani\inst{3}, M.~A. Dupret\inst{4}, S. Mathis\inst{5,1}, \and M. Grosjean\inst{4}}

\institute{
 LESIA, Observatoire de Paris, PSL Research University, CNRS, Universit\'e Pierre et Marie Curie,
 Universit\'e Denis Diderot,  92195 Meudon, France
\and
Institut d'Astrophysique Spatiale, CNRS, Universit\'e Paris XI,
   91405 Orsay Cedex, France
\and
Stellar Astrophysics Centre, Department of Physics and Astronomy, Aarhus University, 
Ny Munkegade 120, DK-8000 Aarhus C, Denmark
\and
Institut d'Astrophysique et de G\'eophysique, Universit\'e de Li\`ege, All\'ee du 6 Ao\^ut 17-B 4000 Li\`ege, Belgium 
\and
Laboratoire AIM Paris-Saclay, CEA/DSM-CNRS-Universit\'e
Paris Diderot; IRFU /SAp, Centre de Saclay, 91191 Gif-sur-Yvette Cedex, France
}

   \offprints{K. Belkacem}
   \mail{kevin.belkacem@obspm.fr}
   \date{\today}

  \authorrunning{Belkacem}

   \abstract{ 
   The detection of mixed modes in subgiants and red giants by the CoRoT and \emph{Kepler} space-borne missions allows us to investigate the internal structure of evolved low-mass stars, from the end of the main sequence to the central helium-burning phase. In particular, the measurement of the mean core rotation rate as a function of the evolution places stringent constraints on the physical mechanisms responsible for the angular momentum redistribution in stars. It showed that the current stellar evolution codes including the modelling of rotation fail to reproduce the observations. An additional physical process that efficiently extracts angular momentum from the core is thus necessary. 
   
   Our aim is to assess the ability of mixed modes to do this. To this end, we developed a formalism that provides a modelling of the wave fluxes in both the mean angular momentum and the mean energy equations in a companion paper. In this article, mode amplitudes are modelled based on recent asteroseismic observations,  and a quantitative estimate of the angular momentum transfer is obtained. 
   This is performed for a benchmark model of 1.3 $M_{\odot}$ at three evolutionary stages, representative of the evolved pulsating stars observed by CoRoT and Kepler. 
   
   We show that mixed modes extract angular momentum from the innermost regions of subgiants and red giants. However, this transport of angular momentum from the core is unlikely to counterbalance the effect of the core contraction in subgiants and early red giants. In contrast, for more evolved red giants, mixed modes are found efficient enough to balance and exceed the effect of the core contraction, in particular in the hydrogen-burning shell. Our results thus indicate that mixed modes are a promising candidate to explain the observed spin-down of the core of evolved red giants, but that an other mechanism is to be invoked for subgiants and early red giants.
  }


   \maketitle

\section{Introduction}
\label{intro}

The CoRoT \citep{Baglin2006a,Baglin2006b,Michel2008} and \emph{Kepler} \citep{Borucki2010} space-borne missions provided a wealth of observed stars exhibiting solar-like oscillations, from the main-sequence to the red-giant phases \citep[see][for a review]{Chaplin2013}. A large number of those stars are low-mass evolved stars showing a rich spectrum of mixed modes, which behave as acoustic modes in the stellar envelope and as gravity modes in the core. They are therefore detectable at the stellar surface while yielding information on the innermost regions \citep[e.g.,][]{Dziembowski1971,Scuflaire1974,Aizenman1977,Dziembowski2001,Dupret2009}. 

While mixed modes allowed unveiling the structure of the red giants core \citep[][]{Bedding2011,Mosser2011,Mosser2014}, the detection of associated rotational splittings enabled measuring the mean core rotation of subgiant and red giant stars  \citep{Beck2012,Deheuvels2012,Mosser2012,Deheuvels2014}. It turns out that the core of low-mass red giant stars slows down during evolution. Since local conservation of angular momentum would imply a  spin-up of the core, these observations indicate a strong transfer of angular momentum from the inner to the outer layers of these stars. However, current stellar evolutionary rotating models that take the transport of angular momentum into account fail to reproduce the observations. Meridional circulation or shear instabilities are not efficient enough to slow down the red-giant cores \citep[e.g.,][]{Eggenberger2012,Marques2013,Ceillier2013}, and recent attempts to include the effect of propagative internal gravity waves \citep{Fuller2014} or magnetic fields using the Taylor-Spruit dynamo formalism \citep{Cantiello2014} failed to solve the problem. Nonetheless, for subgiants and early red giants, 
\cite{Rudiger2015}  have recently shown  that magneto-rotational instabilities of a toroidal magnetic field could explain the angular momentum redistribution in subgiants and early red giants \citep[see also][]{Maeder2014}. 

Hence, an additional physical mechanism that efficiently extracts angular momentum from the core of red giants seems to be needed. Non-radial modes in subgiants and red giants are potential candidates. Indeed, prograde and retrograde mixed modes are differentially damped in the presence of rotation, allowing a net transport of angular momentum \citep[e.g.,][]{Ando1986,Lee1993}. Our aim is thus to investigate the ability of mixed modes to transfer angular momentum from the inner radiative interior to the outer convective layers.

In the first article of this series (hereafter Paper I), we used the transformed Eulerian mean (TEM) formalism to account for the effect of wave driving through both the wave momentum flux in the mean angular momentum equation and the wave heat flux in the mean energy equation. The wave field was then modelled using the asymptotic, quasi-adiabatic, and slow-rotation approximations, valid in the dense radiative layers of evolved stars. This allowed us to obtain an explicit expression of the wave fluxes appearing in the mean angular momentum and energy equations. This article, the second of this series, is dedicated to a quantitative determination of the angular momentum transported by mixed modes, which requires the knowledge of mode amplitudes. Recent CoRoT and \emph{Kepler} observations \citep{Baudin2011b,Mosser2012b,Samadi2012} allow us to determine mixed-mode amplitudes and subsequently to determine the efficiency of angular momentum transport by mixed modes.

The paper is organized as follows: Sect.~\ref{formalism} recalls the formalism developed in Paper~I, while Sect.~\ref{models} presents the modelling of the equilibrium models and related oscillations. In Sect.~\ref{amplitudes_modes}, based on the observations, we describe the modelling of mode amplitudes. Finally, the estimate of the angular momentum transport efficiency is presented in Sect.~\ref{results}, and Sect.~\ref{conclusions} is dedicated to discussions and conclusions. 

\section{Theoretical formalism}
\label{formalism}

In this section, we recall the formalism developed in Paper I as well as the related assumptions for both the mean flow and the wave fluxes appearing in the mean equations. 

\subsection{Mean flow equations}
\label{meanflow}

Following Paper I, we adopt the transformed Eulerian mean formalism (TEM). In addition, we assume shellular rotation \citep[see][for an extensive discussion]{Maeder09} so that at dominant order the mean angular momentum equation is
\begin{align}
\label{shellular_omega_r}
&\left< \rho \right> \deriv{\left( r^2 \Omega_0 \right)}{t}   
= -\frac{1}{r^2} \derivp{}{r} \left[ r^2 \left( \mathcal{F}_{\rm circ} +  \mathcal{F}_{\rm shear} + \mathcal{F}_{\rm waves} \right) \right] \,, 
\end{align}
where $\Omega_0$ is the rotation angular frequency that depends only on the radius, $\rho$ is the density, the overbar and the symbol $\left< \right>$ denotes the horizontal average (\emph{i.e.} azimuthal and meridional). The fluxes are defined by
\begin{align}
\label{shellular_omega_r_advection}
\mathcal{F}_{\rm circ} &=  -\frac{1}{5} \left< \rho \right> r^2 \Omega_0 \, U_2^\dag \,  , \\
\label{shellular_omega_r_shear}
\mathcal{F}_{\rm shear} &=  -\left< \rho \right> \nu\ind{v} r^2 \derivp{\Omega_0}{r}\,  ,  \\
\label{shellular_omega_r_wave}
\mathcal{F}_{\rm waves} &=  \left< \rho \right>\left< \varpi \left[ \overline{\vt_\phi^\prime \, \vt_r^\prime} + 2 \cos \theta \,  \Omega_0 \, \overline{\vt_\theta^\prime s^\prime} \left(\deriv{\left< s\right>}{r}\right)^{-1} \right] \right> \, ,
\end{align}
where $\varpi = r \sin \theta$, the overbar is the azimuthal average, $U_2^\dag$ is the residual meridional velocity, $\nu_v$ is the radial eddy viscosity, $s$ is the specific entropy, and the prime denotes perturbations associated with the non-radial oscillations, so that $\vt_\phi^\prime$, $\vt_r^\prime$, $\vt_\theta^\prime$ are the azimuthal, radial, and meridional component of the wave velocity field, and $s^\prime$ the wave Eulerian perturbation of entropy.  We have also introduced the Lagrangian derivative ${\rm d} / {\rm d}t =  \partial / \partial t +  \dot{r} \, \partial / \partial r$. 

Under the same assumptions as for \eq{shellular_omega_r}, the mean entropy equation is  
\begin{align}
\label{energy_shellular_vertical}
 \left< \rho \right> \, \deriv{ \left< s \right>}{t}  = -\frac{1}{r^2} \derivp{}{r} \left< r^2 \,\mathcal{S}\right> + \left< Q \right> \, 
\end{align}
with 
\begin{align}
\label{energy_shellular_vertical_flux}
\mathcal{S} = \left< \rho\right>\,  \overline{s^\prime \, \vt_r^\prime} \, , 
\end{align}
and
\begin{align}
 \left< T\right> \left< Q \right> = \left< \rho \varepsilon \right> + \frac{ 1}{r^2} \frac{\partial}{\partial r}\left(r^2 \left< \chi \right> \frac{\partial \left<T\right>}{\partial r}\right)\, ,
\end{align}
where $T$ is the temperature, $\varepsilon$ is the nuclear energy generation rate and $\chi$ is the thermal  conductivity (see, e.g., Sect. 6 of \citealt{Mathis2004}, and also \citealt{MaederZahn1998}). This expresses energy conservation on level surfaces.

To quantify the effect of the waves on the rotation profile, we further need to provide
a modelling of the wave field\footnote{In this article, by \emph{\textup{wave}} we denote global mixed modes.}. 

\subsection{Wave flux modelling}
\label{wavefield}

Our aim is to determine the angular momentum redistribution induced by mixed modes in the radiative region of low-mass evolved stars. Therefore, the wave fluxes in Eqs.~(\ref{shellular_omega_r}) and (\ref{energy_shellular_vertical}) must be modelled.  This has been done in Paper I, so that for the momentum equation we have
\begin{align}
\label{final_F_waves}
&-\frac{1}{r^2} \derivp{}{r} \left(r^2 \mathcal{F}_{\rm waves}\right) = \\
&\sum_{\ell,m} a_{\ell,m}^{2} \left( \mathcal{A}_{\ell}^m \; \derivp{^2\left(r^2 \Omega_0 \right)}{r^2} +\mathcal{B}_{\ell}^m \; \derivp{\left(r^2 \Omega_0 \right)}{r} + \mathcal{C}_{\ell}^m \; \Omega_0 + m \hat \sigma \mathcal{D}_{\ell}^m \right) \nonumber \, , 
\end{align}
where the coefficients $\mathcal{A}_{\ell}^m, \mathcal{B}_{\ell}^m, \mathcal{C}_{\ell}^m, \mathcal{D}_{\ell}^m$ are given by Eqs.~(A.25) to (A.28) of Paper I, and $\hat \sigma = \sigma_R + m\Omega_0$ (with $\sigma_R$ the modal frequency). The amplitude $a_{\ell,m}$ corresponds to the amplitude of a mode of a given angular degree, $\ell$, and azimuthal order, $m$. We note that the amplitude is considered statistically constant in time since we deal with solar-like oscillations that result from a balance between mode driving and damping \citep[see][for details]{Samadi11}. 

For the entropy equation, we obtained 
\begin{align}
\label{final_T_2}
\frac{1}{r^2} \derivp{}{r} \left< r^2 \,\mathcal{S} \right> = \sum_{\ell,m} \frac{a_{\ell,m}^{2}}{2 r^2} \derivp{}{r} \left( r^2 \, \rho \, \alpha \deriv{s}{r} k_r^2 \, \vert \xi_r^{\ell,m}\vert^2  \right) \, ,
\end{align}
where 
\begin{align}
&\alpha = -\frac{L}{4 \pi r^2 \rho T} \left( \frac{\nabla_{\rm ad}}{\nabla}-1 \right) \, \left(\deriv{s}{r}\right)^{-1}\nonumber \\
&k_r^2 \simeq \frac{\ell(\ell+1)}{r^2} \, \left(\frac{N^2}{\sigma_R^2}-1\right) \, , 
\end{align}
with $L$ the luminosity, $N$  the buoyancy frequency, $\nabla$ and $\nabla_{\rm ad}$  the actual and adiabatic temperature gradients, respectively, and $\xi_r^{\ell,m}$ the radial component of the eigenfunction for a mode of angular degree $\ell$ and azimuthal order $m$. The overbars and angular brackets have been dropped for ease of notation. 

We also recall that in the course of deriving Eqs.~(\ref{final_F_waves}) and (\ref{final_T_2}) several simplifying approximations were performed, namely:  
\begin{enumerate}
\item \emph{The quasi-adiabatic approach:} It consists of neglecting the difference between adiabatic and non-adiabatic eigenfunctions in the full wave equations. This approximation is valid provided the local thermal timescale is much longer than the modal period. This is justified in the radiative region of evolved low-mass stars. 
\item \emph{The low-rotation limit:} We consider the low-rotation limit in the radiative regions or, more precisely,   $\sigma_R \gg \Omega_0$. This is justified by recent inferences of the rotation rate in the core of subgiants \citep{Deheuvels2012,Deheuvels2014} and red giants \citep{Mosser2012} using seismic constraints from \emph{Kepler}. 
\item \emph{The asymptotic limit:} We used an asymptotic description for gravity modes \citep[e.g.,][]{Dziembowski2001,Godart2009}, which is valid for mixed modes in the inner radiative region of subgiants and red giants \citep[e.g.,][]{Goupil2013}. 
\end{enumerate}
Consequently, our formalism is valid in the dense radiative core of evolved slowly rotating stars. 

\section{Computing selected stellar models}
\label{models}

\begin{figure}
\begin{center}
\includegraphics[width=9cm]{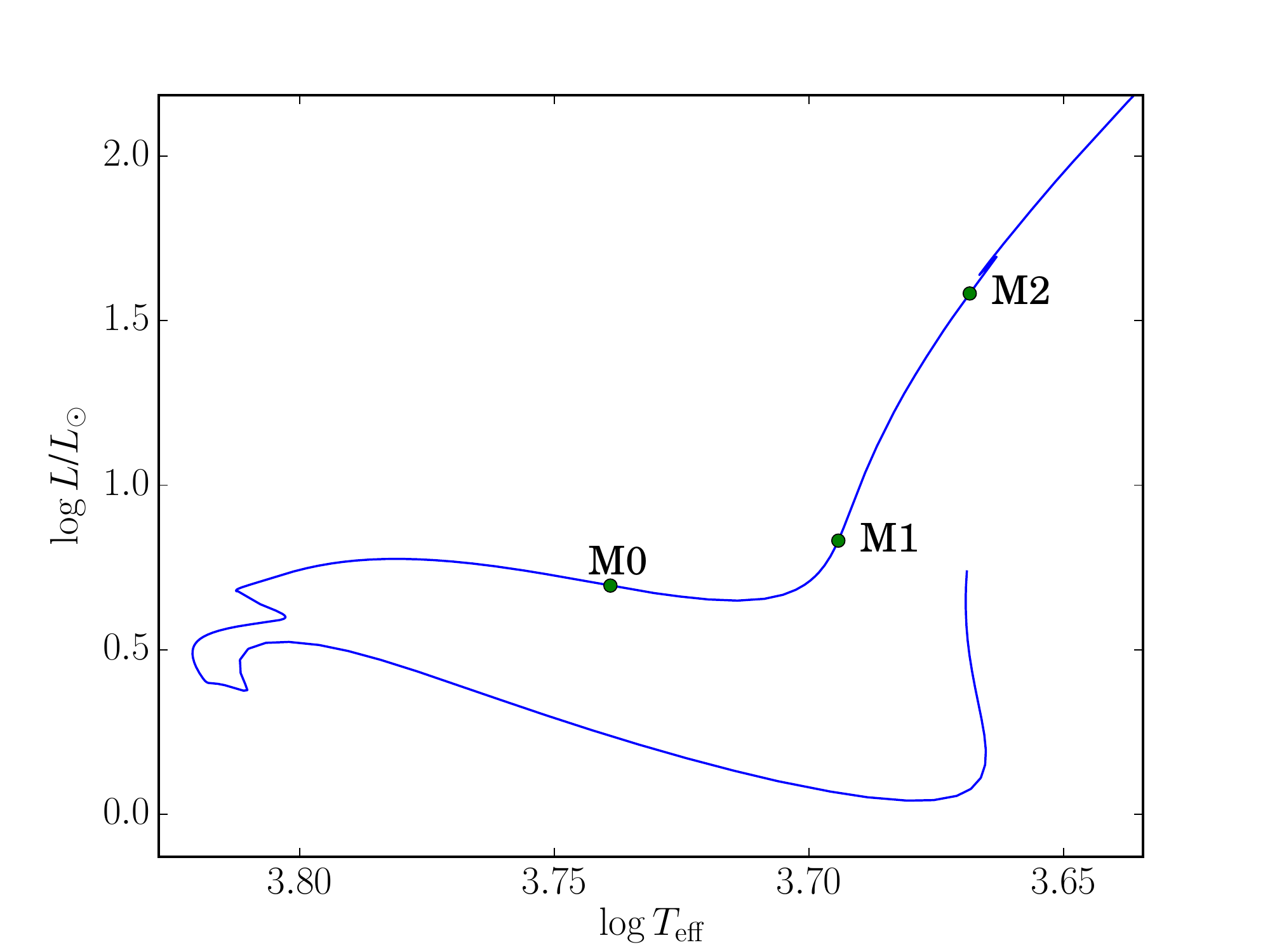}
\caption{Evolutionary track on the Hertzsprung-Russell (HR) diagram of a $1.3 M_{\odot}$ model, showing the location of the selected models (see Sect.~\ref{models} for details). }
\label{fig:hr}
\end{center}
\end{figure}

We have computed an evolutionary sequence of a benchmark model of a 1.3 M$_{\odot}$ evolved low-mass star representative of CoRoT and \emph{Kepler} observations. We selected three $M=1.3 M_\odot$ models at three post-main sequence stages of this evolutionary sequence, as discussed below. 
The location of the models in the Hertzsprung-Russell (HR) diagram is shown in Fig.~\ref{fig:hr}, and their characteristics are summarised in Table~\ref{tab:mods}. 

\begin{table}
   \centering
    \caption{Characteristics of the selected models as described in Sect.~\ref{models}. Model M0 is a subgiant, model M1 is at the base of the red-giant branch, and model M2 is a red giant. Here, $\Delta \nu$ is the large separation and $\nu_{\rm max}$ is the frequency at the maximum height of the oscillation power spectrum. }
  \begin{tabular}{lccccc} 
      \hline
      Model  & Age (Myr)  & $R/R_{\odot}$ & $L/L_{\odot}$  & $\Delta \nu$ ($\mu$Hz) & $\nu_{\rm max}$ ($\mu$Hz)\\
      \hline
      M0      & 4401 & 2.47 & 4.95 & 39.5 & 666 \\
      M1      & 4590 & 3.55 & 6.78 & 22.9 & 339 \\
      M2      & 4791 & 9.51 & 38.3 & 5.24 & 48.8\\
      \hline
   \end{tabular}
   \label{tab:mods}
\end{table}

\subsection{Equilibrium models and related eigenfunctions and eigenfrequencies}

We used the stellar evolution code CESTAM  \citep{Marques2013} to compute the equilibrium models. The atmosphere was computed assuming a grey Eddington approximation. Convection was included according to \cite{CGM1996}, with a mixing-length parameter $\alpha=0.67$.  The initial chemical composition follows \cite{Asplund05}, with a helium mass fraction of $Y=0.261$ and a metallicity of  $Z=0.0138$. We used the OPAL equation of state
\citep{Rogers96} and opacities \citep{Iglesias96}, 
complemented, at $T < 10^4$ K, by the \cite{Alexander94} opacities. 
We adopted the NACRE nuclear reaction rates from \cite{Angulo99} except for the $\element[][14]{N}  + \element{p}$ reaction, where we used the reaction rates given in \cite{Imbriani04}.  

Finally, we used the ADIPLS code to compute adiabatic oscillations \citep{JCD08,JCD11}, that is,\emph{} to obtain the  eigenfunctions and eigenfrequencies.

\subsection{Rotation profiles}
\label{rotation_profiles}

We consider that rotation has negligible effects on the equilibrium structure of the star and that it can be considered as a perturbation for the oscillations. This approximation is consistent with the formalism developed in Paper I. 

\begin{figure}[t]
\begin{center}
\includegraphics[width=9.2cm]{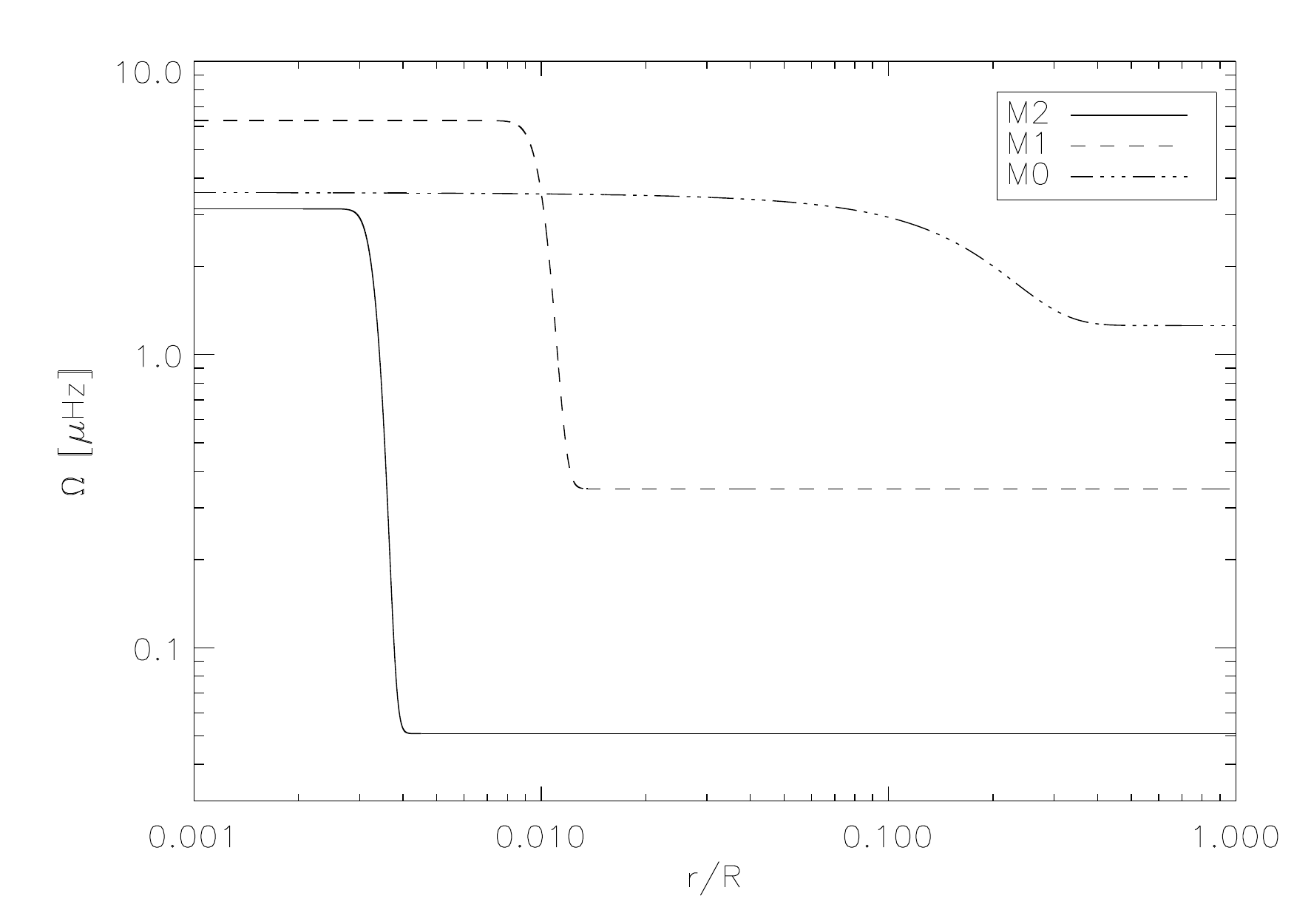}
\caption{Synthetic rotation profiles, computed as described in Sect.~\ref{rotation_profiles}, for models M0, M1, and M2 as a function of the radius normalised by the total radius. We used logarithmic scales.  }
\label{plot_rotations}
\end{center}
\end{figure}

The rotation profiles are therefore determined  by computing synthetic profiles that match the observations. The synthetic profiles should have a core that rotates faster than the envelope, with a transition region located at the H-burning shell in the case of red giant stars (models M1 and M2). The width of the transition region is related to the width of the H-burning shell. We consequently used the following profile 
\begin{align}
\Omega(r) = \Omega_S + \frac{\Omega_C - \Omega_S}{2} \left[1 + {\rm erf} \left(\frac{r_t - r}{w} \right)\right],
\label{eq:synthprof}
\end{align}
where $\Omega_S$ and $\Omega_C$ are the surface and central angular velocities, respectively, and $r_t$ and $w$ are the radius and width of the burning shell. $\Omega_C$ was chosen to reproduce the mean core rotation rate of evolved low-mass stars as derived from \emph{Kepler} observations \cite[see][]{Mosser2012,Deheuvels2014}, and the surface rotation rate ($\Omega_S$) was obtained so as to conserve the total angular momentum from the subgiant stage (model M0).

For the subgiant model (model M0), we also used Eq.~(\ref{eq:synthprof}), but with both $r_t$ and $w$ given by the radius of the central convective zone at the end of the main sequence (just before it disappears). In this case, the central  and surface rotation ($\Omega_C$ and $\Omega_S$, respectively) were chosen to reproduce the results of \cite{Deheuvels2012}. The values of the parameters for models M0, M1, and M2 are summarised in Table~\ref{tab:rotation}, and the resulting rotation profiles are displayed in Fig.~\ref{plot_rotations}.

\begin{table}
   \centering
    \caption{Parameters of the synthetic profiles as described in Sect.~\ref{rotation_profiles} (see also Eq.~\ref{eq:synthprof}). }
  \begin{tabular}{lcccc} 
      \hline
      Model  & $\Omega_C / 2\pi$ (nHz)   & $\Omega_S / 2\pi$ (nHz) & $ r_t/R_{\rm star}$  & $w/R_{\rm star}$ \\
      \hline
      M0      & 600 & 200 & 0.1452 & $1.452 \; 10^{-1}$\\
      M1      & 1000 & 55 & $1.009 \; 10^{-2}$ & $1.211 \; 10^{-3}$ \\
      M2      & 500 & $8.1 $ & $3.316 \; 10^{-3}$ & $3.015 \; 10^{-4}$\\
      \hline
   \end{tabular}
   \label{tab:rotation}
\end{table}

\section{Modelling mode amplitudes}
\label{amplitudes_modes}

There are mainly two approaches  to compute mode amplitudes. The first  is based on a full non-adiabatic computation including a time-dependent treatment of convection. This procedure is time-consuming, however, and still suffers from theoretical uncertainties  \citep[see][for details]{Dupret2009,Grosjean2014}. 
The second is based on recent CoRoT and \emph{Kepler} observations that allowed us to measure the amplitudes of solar-like modes in a wide variety of stars in different evolutionary stages. This wealth of observations made it possible to establish scaling relations that provide mode amplitudes versus global stellar parameters \citep[e.g.,][]{Mosser2012b,Samadi2012}. In this paper, we adopt the latter approach.  

\begin{figure}[t]
\begin{center}
\includegraphics[width=9.2cm]{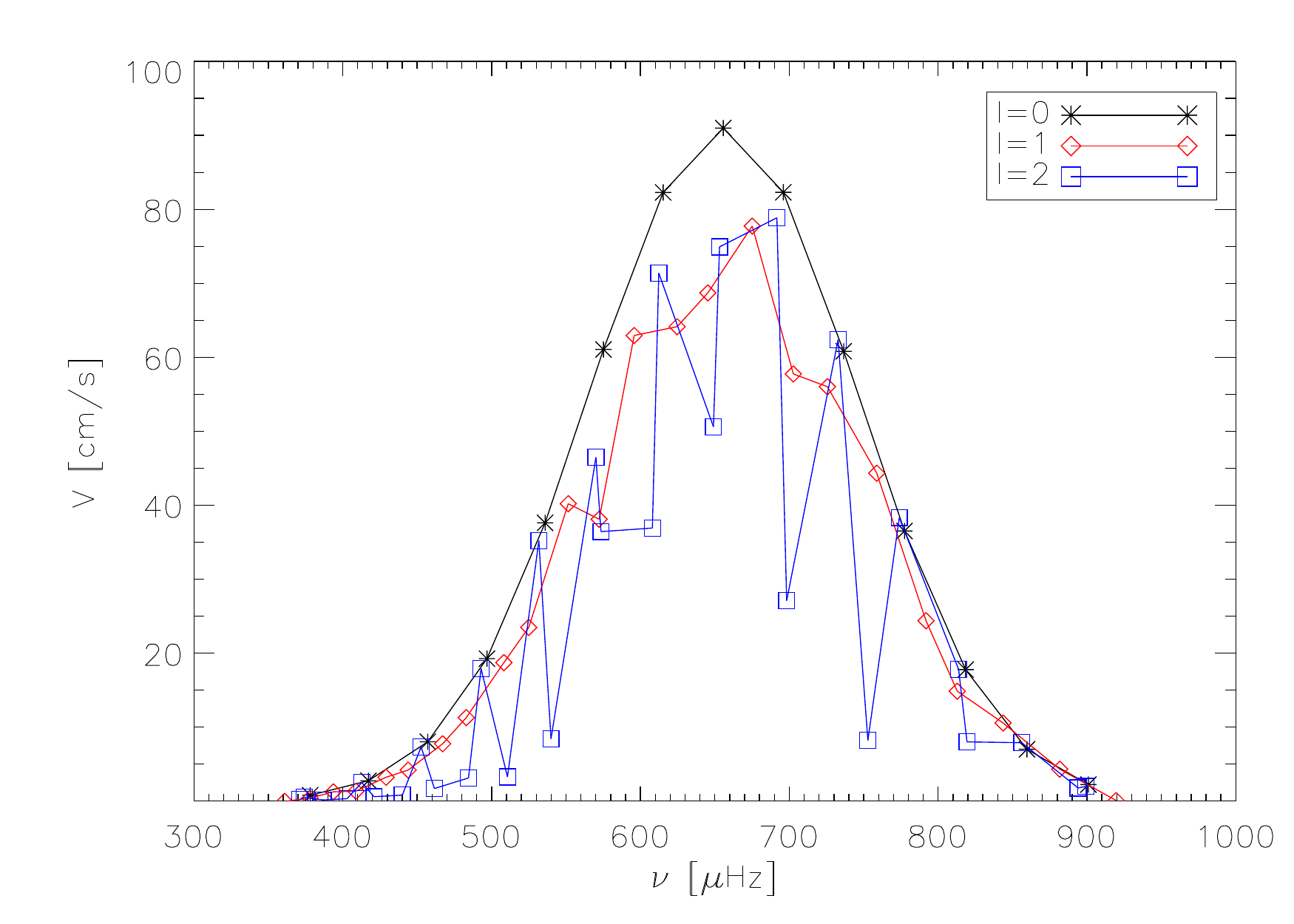}
\includegraphics[width=9.2cm]{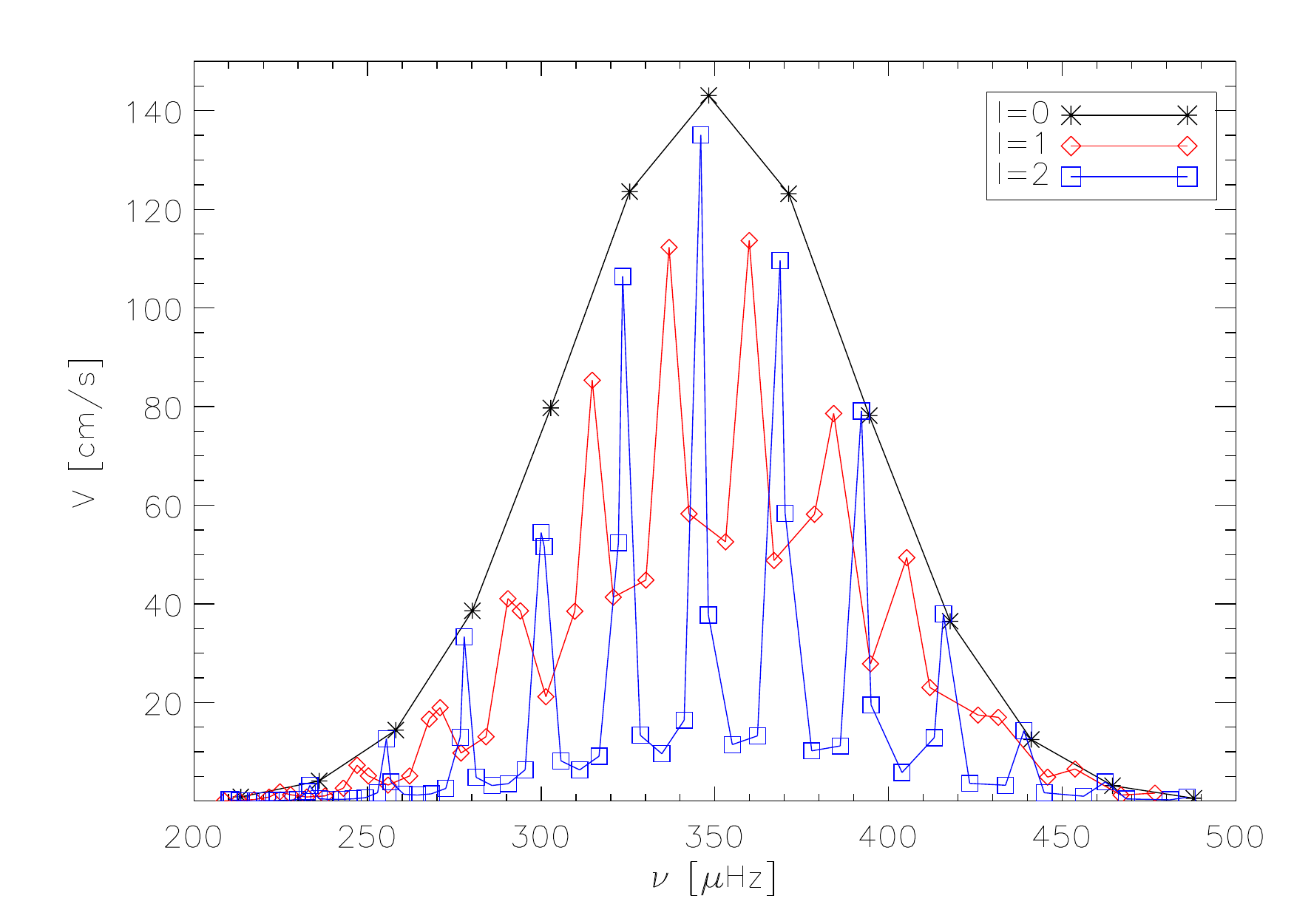}
\includegraphics[width=9.2cm]{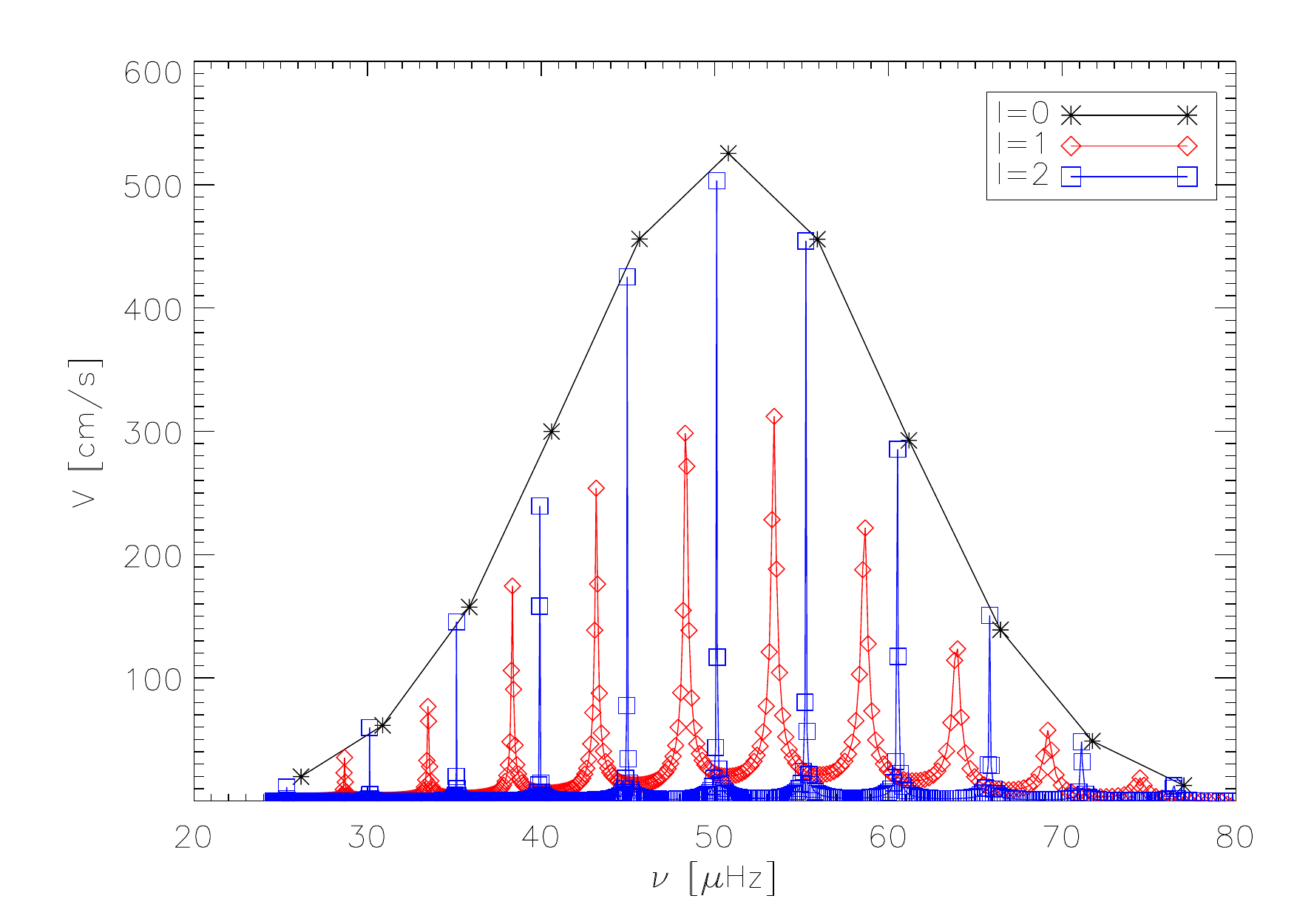}
\caption{Mode amplitudes versus mode frequencies for models M0 (top panel), M1 (middle panel), and M2 (bottom panel). The amplitudes are computed as described in Sect.~\ref{amplitudes_modes}.
\label{plot_amplitudes}}
\end{center}
\end{figure}

\subsection{Radial mode amplitudes}
\label{radial}

We first express the amplitudes of radial modes. The surface velocity of the radial modes, at the frequency of maximum height ($\nu_{\rm max}$) in the oscillation power spectrum, follows the scaling relation \cite[see][for details]{Belkacem2013}
\begin{align}
\frac{V_{\ell=0}\left(\nu = \nu_{\rm max} \right) }{V_{\ell=0, \odot} \left(\nu = \nu_{\rm max} \right)} \approx \left(\frac{T_{\rm eff}}{T_{\odot,\rm eff}}\right)^{-1.77} \, \left(\frac{\nu_{\rm max}}{\nu_{\odot,\rm max}}\right)^{-1.15} \, \left(\frac{\Delta \nu}{\Delta \nu_{\odot}}\right)^{0.65} \, , 
\end{align}
where $T_{\rm eff}$ is the effective temperature, $\Delta \nu$ is the large separation, and the $\odot$ symbols stand for the solar values. Here we adopt the values $T_{\odot,\rm eff} = 5777$ K, $\nu_{\odot,\rm max} = 3050 \, \mu$Hz, and $\Delta \nu_{\odot} = 135 \, \mu$Hz. 

To account for the frequency dependence of radial mode amplitudes, we follow the work of \cite{Mosser10} , who used a Gaussian envelope centred on $\nu_{\rm max}$ with a full-width at half-maximum $\delta \nu_{\rm env}$, that is, \emph{} 
\begin{align}
V_{\ell=0}^2\left(\nu \right) = V_{\ell=0}^2 \left(\nu = \nu_{\rm max} \right) \; e^{-\left(\nu-\nu_{\rm max} \right)^2 / \left(2 \sigma^2\right) } \, , 
\end{align}
with $\sigma = \delta \nu_{\rm env} / (2 \sqrt{2 \ln 2})$. 
Based on \emph{Kepler} observations, we  take $\delta \nu_{\rm env}=150 \,\mu$Hz for M0,  $\delta \nu_{\rm env}=70 \,\mu$Hz for M1, and $\delta \nu_{\rm env}=30 \,\mu$Hz for M2  for the following. 

\subsection{Non-radial mode amplitudes}
\label{nonradial}

The next step is to consider the amplitudes of non-radial modes. The ratio between radial and non-radial mode amplitudes of comparable frequencies can be written  as \cite[see][for details]{Benomar2014}
\begin{align}
\label{amplitudes_othman}
\frac{V_\ell^2 \left(\nu \right)}{V_{\ell=0}^2 \left(\nu \right)} \approx \frac{\mathcal{M}_{\ell=0}^2 }{\mathcal{M}_{\ell}^2} \,  \frac{\Gamma_{\ell=0}}{\Gamma_{\ell}} \, , 
\end{align}
where $\mathcal{M}_\ell= \int \vec \xi^2 {\rm d}m / \xi^2(R)$ is the mode mass, $\vec \xi$ is the eigenfunction, ${m}$ is the mass profile, $R$ is the total radius, and $\Gamma_{\ell}$ is the mode linewidth that is related to the mode damping  rate $\eta_{\ell}$ by $\Gamma_{\ell} = \eta_{\ell} / \pi$. 

For high angular degrees, the radiative work of the oscillations in the innermost regions dominantes the work done in the uppermost layers \citep[see][for details]{Grosjean2014}. It follows that the mode amplitudes of high angular degrees become very small due to strong radiative damping. We thus define a cut-off angular degree, $\ell_{\rm max}$, below which we assume that the work of non-radial modes equals the work  of radial modes  
\begin{align}
\label{equality_works}
\Gamma_{\ell=0} \mathcal{M}_{\ell=0} = \mathcal{M}_{\ell \leq \ell_{\rm max}} \Gamma_{\ell \leq \ell_{\rm max} } \, .
\end{align}
where the product between the mode mass and the mode damping rate equals the work integral \cite[e.g.,][]{Dupret2009,Benomar2014,Grosjean2014}. 
Now, using \eq{equality_works} together with \eq{amplitudes_othman} leads to 
\begin{align}
\label{amplitudes_final}
\frac{V_{\ell \leq \ell_{\rm max}}^2 \left(\nu \right)}{V_{\ell=0}^2 \left(\nu \right)} \approx \frac{\mathcal{M}_{\ell=0} }{\mathcal{M}_{\ell \leq \ell_{\rm max}}} \, .  
\end{align}
For $\ell > \ell_{\rm max}$, we assume that the radiative damping in the innermost region becomes dominant over the damping of the upper layers so that mode amplitudes become negligible. 

The cut-off angular degree is obtained by assuming that the work done over an oscillation period through the effect of radiative losses in the innermost layers becomes of the same order of magnitude as the work done in the uppermost layers. We obtain $\ell_{\rm max}=6$ for M0, $\ell_{\rm max}=4$ for M1, and $\ell_{\rm max}=2$ for M2 (see Appendix~\ref{cutoff} for details). Note that this approach for the modelling of mode amplitudes provides a lower limit to the amount of angular momentum transported by mixed modes since we neglect the contribution of angular degrees higher than $\ell_{\rm max}$. However, this is enough to determine whether mixed modes are able to slow down the rotating core of red giants. 

Finally, we assume that the relation between the surface mode velocity and the mode amplitude is given by $V_\ell^2 = a_{\ell,m}^2 \sigma_R^2 \vert \vec \xi_\ell \vert^2 / 2$ \citep[see][]{Samadi11}.  The results obtained for mode amplitudes of angular degrees $\ell=\{0,1,2\}$ are presented in Fig.~\ref{plot_amplitudes} for the three models M0, M1, and M2,  described in Sect.~\ref{models}. 

\section{Quantitative estimate of angular momentum extracted by mixed modes}
\label{results}

Our aim is to estimate the rate of angular momentum transported by mixed modes for the benchmark models M0, M1, and M2. To this end, as in Paper I, we introduce the notation
\begin{align}
\label{Jdot}
\dot{J} = -\frac{1}{r^2} \derivp{}{r} \left( r^2 \mathcal{F}_{\rm waves} \right) \, ,
\end{align}
where the right-hand side is given by \eq{final_F_waves}. Using the approach presented in Sect.~\ref{amplitudes_modes} to derive the amplitudes, 
we compute the wave fluxes in the right-hand side of Eqs.~(\ref{shellular_omega_r}) and (\ref{energy_shellular_vertical}) using Eqs.~(\ref{final_F_waves}) and (\ref{final_T_2}). 

\begin{figure}[t]
\begin{center}
\includegraphics[width=9.2cm]{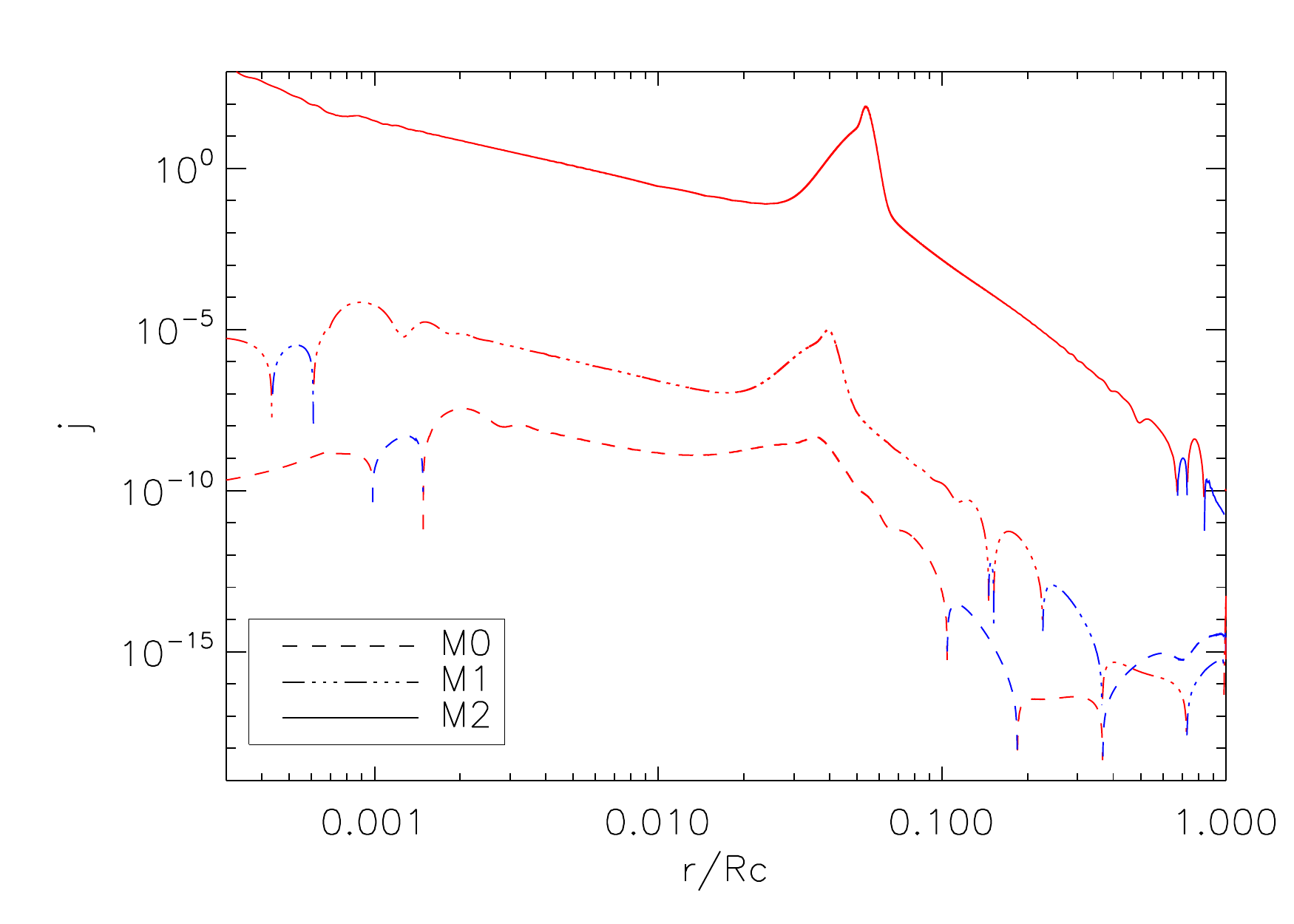}
\caption{Rate of temporal variation of the mean specific angular momentum ($\dot{J}$ as defined by Eq.~\ref{Jdot}) as a function of the star radius normalised by the radius of the base of the convective envelope. Red  corresponds to $\dot{J}< 0$ and blue to $\dot{J} > 0$. }
 \label{plot_fluxes}
\end{center}
\end{figure}

The contribution of the mixed modes to the mean angular momentum equation is shown in Fig.~\ref{plot_fluxes}  for models M0, M1, and M2. The main effect of mixed modes is to slow down the core of the stars in the region near the maximum of the buoyancy frequency that also corresponds to the edge of the fast rotating core. For the red-giant models, this region is the hydrogen-burning shell. 
The decrease shown towards the upper radiative layers is the result of both a lower rotation rate and a lower value of the buoyancy frequency. 
This behaviour is similar in the three models. 
In addition, the amount of angular momentum extracted by mixed modes  hardly depends on the rotation gradient between the core and the envelope (the sensitivity of the results to the parameter $w$ in \eq{eq:synthprof} remains weak). 

The amount of angular momentum extracted by mixed modes increases with the evolutionary stage of the star. This effect is the result of several factors. First, as shown in Fig.~\ref{plot_amplitudes}, mode amplitudes increase from models M0 to M2 and thus more energy is available to transport angular momentum. Moreover, the buoyancy frequency also increases and so does the radial wave number in the gravity-mode cavity. As the energy exchanges between modes and the background is proportional to the radial wave number, the amount of angular momentum extracted also increases. Finally, the number of mixed modes between two p-dominated modes significantly increases between M0 and M2. 

Furthermore, the angular momentum is mainly carried away by g-dominated modes of the highest angular degrees. Indeed, these modes are very efficiently trapped in the core and undergo a larger radiative damping (due to the larger radial wave number). The predominance of high angular degree modes implies that $\dot{J}$ is sensitive to our determination of $\ell_{\rm max}$. As mentioned before, the estimate of the amount of angular momentum extracted in the radiative regions is thus a lower limit. In other words, adding the contribution of angular degrees higher than $\ell_{\rm max}$ would strengthen these trends. 

To proceed and  assess the efficiency of the transport of angular momentum for our benchmark models, we 
compare two timescales, namely: 
\begin{itemize}
\item the timescale associated with the efficiency of the transport of angular momentum by mixed modes, defined as
\begin{align}
\label{time_modes}
T_m^{-1} = \left\vert \frac{\dot{J}}{\rho r^2 \Omega_0} \right\vert \, ,
\end{align}
\item the timescale associated with the contraction of the star, defined as
\begin{align}
\label{time_spin}
T_c^{-1} =\left\vert -\frac{1}{\rho r^4 \Omega_0} \derivp{}{r} \left(\rho r^4 \Omega_0 \, \dot{r}\right) \right\vert  \approx \left\vert - \frac{\dot{r}}{r} \right\vert  \, ,
\end{align}
where $\dot{r}={\rm d} r / {\rm d} t$.  
\end{itemize} 

\begin{figure}[t]
\begin{center}
\includegraphics[width=9.2cm]{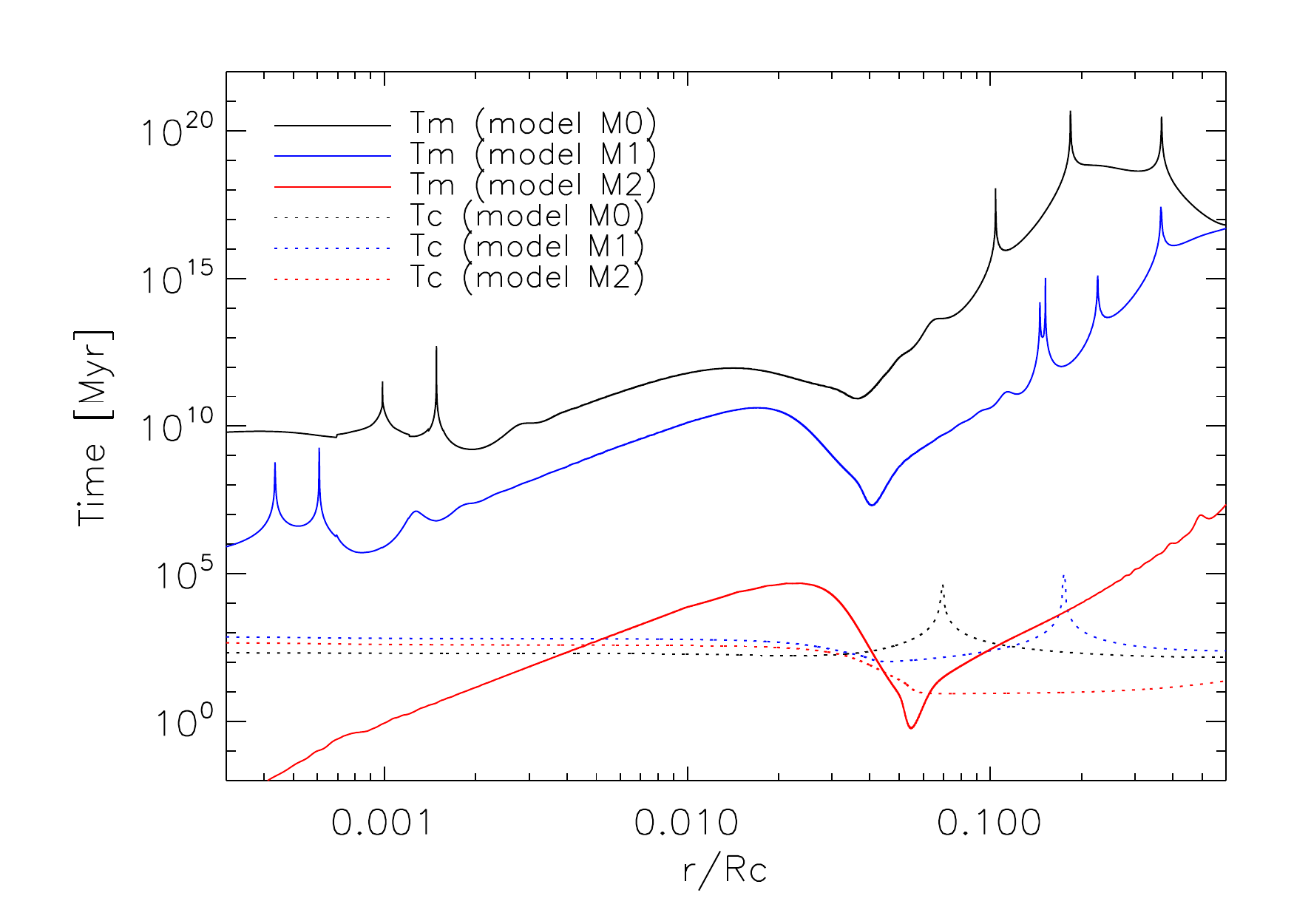}
\includegraphics[width=9.2cm]{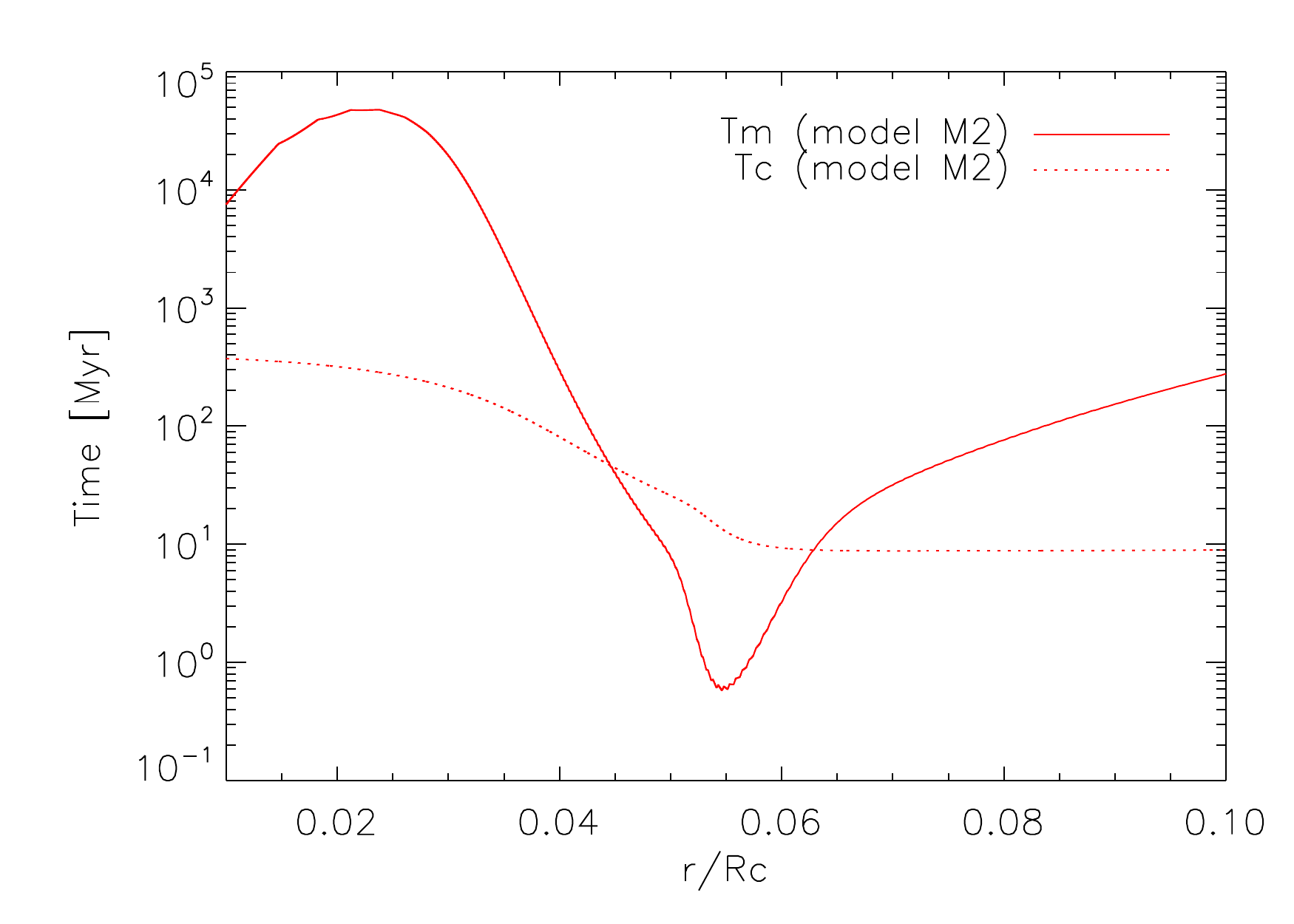}
\caption{{\bf Top panel:} Timescales versus normalised radius (\emph{i.e.}, normalised by the radius of the base of the convective envelope) for models M0, M1, and M2. The solid lines correspond to the timescale associated with the transport of angular momentum by mixed modes (see Eq.~\ref{time_modes}) and the dotted lines correspond to the timescale associated with the contraction of the star (see Eq.~\ref{time_spin}). 
{\bf Bottom panel:} Same as the top panel, except that only a zoom for model M2 is shown. 
\label{plots_timescales}}
\end{center}
\end{figure}

The two timescales are shown in Fig.~\ref{plots_timescales} (top panel). In all the radiative regions of models M0 and M1, the extraction of angular momentum remains negligible compared to the star contraction. In contrast, for model M2, the timescale of angular momentum extraction is of the same order of magnitude as  the contraction timescale and even lower in the hydrogen-shell-burning region, for which the contraction is maximal (see Fig.~\ref{plots_timescales}, bottom panel). However, in the very centre or in the upper radiative region, the mixed mode extraction of angular momentum is  again negligible. 

Therefore, we conclude that the extraction of angular momentum by mixed modes is negligible in subgiants and early red giants,
but becomes strong  in the hydrogen-burning shell in stars higher on the red-giant branch. In these cases, mixed modes are able to counterbalance the spin-up due to the star contraction and can thus enforce a spin-down in those layers. We note, however, that the exact location in the HR diagram of the transition between inefficient and efficient extraction by mixed modes is likely to depend on the mass and internal physics of the models. 

\section{Concluding remarks}
\label{conclusions}

Based on the theoretical formalism developed in paper I, we computed the amount of angular momentum extracted by mixed modes in evolved low-mass stars. To this end, we considered $1.3 M_\odot$ benchmark models of a subgiant (model M0), an early red giant (model M1), and a more evolved red giant  (model M2) for which we computed realistic mode amplitudes based on recent CoRoT and \emph{Kepler} observations. We found that mixed modes extract angular momentum in the innermost region of subgiants and red giants. This extraction of angular momentum is found negligible for the subgiant and lower RGB models, whereas it is efficient enough to counterbalance the effect of the stellar contraction  for the most evolved model. 

For this evolved star, mixed modes efficiently transport the angular momentum in the hydrogen-burning shell (that corresponds to the maximum of the buoyancy frequency) but becomes less efficient in the very central layers. We note, however, that mixed modes can be considered as an external torque for meridional circulation \citep[e.g.,][]{Mathis2013}. Therefore, such a localized spin-down in the hydrogen-burning shell is likely to feed a strong redistribution of angular momentum in the central layers by meridional circulation. This would nevertheless deserve a quantitative estimate. 

Indeed, evolutionary calculations including the effect of mixed modes coupled with meridional circulation and shear instabilities are the next step for a definitive conclusion on the ability of mixed modes to slow down the cores of evolved red giants. Modelling of mode amplitudes would  require further investigation  
since we assumed that the contribution of high-angular degree modes would be negligible due to their high radiative damping. 
Therefore, we believe that our calculations provide a lower limit. Full non-adiabatic computations would provide more accurate results, however. 

 We conclude that, for subgiants and early red giants, our results indicate that extraction of angular momentum by mixed modes cannot explain the current observations \citep{Deheuvels2014}. Other physical mechanisms such as internal gravity waves \citep{Talon2008,Fuller2014}  or magnetic fields \citep{Rudiger2015} should also be present in these stars in order to extract enough angular momentum from the core. In contrast, as demonstrated in this work, the transport of angular momentum by mixed modes is a serious explanation for the observed spin down of the core of evolved stars on the red-giant branch. At least, given the order of magnitude of the induced angular momentum extraction, it seems difficult to neglect this mechanism in the future. 

Finally, we note that an extension of this work to central helium burning stars would be desirable in the future. 
 Indeed, even if the mode amplitudes are similar for stars on the vertical branch and on the clump for the same $\nu_{\rm max}$ \citep[e.g.,][]{Mosser2012b}, the maximum of buoyancy frequency is lower and shifted toward higher radius for clump stars \citep[e.g.,][]{Montalban2013}. Therefore, we can expect a lower efficiency of the transport by mixed modes but this needs to be definitively quantified.

\begin{acknowledgements}
We acknowledge the ANR (Agence Nationale de la Recherche, France) program IDEE (n$^\circ$ANR-12-BS05-0008) ``Interaction Des \'Etoiles et des Exoplan\`etes'' as well as financial support from "Programme National de Physique Stellaire" (PNPS) of CNRS/INSU, France. RMO aknowledges funding for the Stellar Astrophysics Centre, provided by The Danish National Research Foundation, and funding for the ASTERISK project (ASTERoseismic Investigations with SONG and Kepler) provided by the European Research Council (Grant agreement no.: 267864).
\end{acknowledgements}


\appendix 

\section{Cut-off angular degrees}
\label{cutoff}
   
The cut-off angular degree, $\ell_{\rm max}$ as introduced in Sect.~\ref{nonradial}, can be defined as the angular degree for which the work performed over an oscillation period through the effect of radiative losses in the inner-most layers becomes of the same order of magnitude as for the work performed in the uppermost layers. More precisely, $\ell_{\rm max}$ is obtained at $\nu=\nu_{\rm max}$ through the relation 
\begin{align}
\label{def_lmax}
 \left( \eta_{\ell_{\rm max}} \mathcal{M}_{\ell_{\rm max}} \right)_{\rm upper\;  layers} \approx  \left( \eta_{\ell_{\rm max}} \mathcal{M}_{\ell_{\rm max}} \right)_{\rm inner\;  layers} \, .
\end{align}
The left-hand side of \eq{def_lmax} can be identified with a radial mode of comparable frequency for which the mode mass is computed as described in Sect.~\ref{models} and the associated damping rate is obtained using the scaling relation as provided by \cite{Belkacem2012} \citep[see also][]{Baudin2011b,Appourchaux2012}. It gives $\eta \,  (\ell=0,\nu=\nu_{\rm max}) = 2.2 \, \mu$Hz for M0,  $\eta \,  (\ell=0,\nu=\nu_{\rm max}) = 0.95 \, \mu$Hz for M1, and $\eta \,  (\ell=0,\nu=\nu_{\rm max}) = 0.06 \, \mu$Hz for M2.

The right-hand side of \eq{def_lmax} can be identified with a gravity dominated mode for which the mode mass is computed following Sect.~\ref{models} and the damping rate is obtained using the asymptotic relation \citep[e.g.][]{Dziembowski2001,Godart2009} 
\begin{equation}
\label{eta_asymp}
\eta \left(\ell_{\rm max},\nu \right) = \frac{\left[ \ell_{\rm max} (\ell_{\rm max} +1) \right]^{3/2}}{8 \pi \sigma_R^3 \int_0^{R_c} k_r \, {\rm d}r} 
\int_{0}^{R_c} \frac{\nabla_{\rm ad} - \nabla}{\nabla} \frac{\nabla_{\rm ad} N g L}{P r^5} \;{\rm d}r  \, .
\end{equation}
where $P$ is the pressure, $L$ is the luminosity, $R_c$ is the base of the convective region, $\sigma_R = 2 \pi \nu$, $\nabla$ is the temperature gradient and $\nabla_{\rm ad}$ its adiabatic counter-part. 

Finally, using Eqs.~(\ref{def_lmax}), and (\ref{eta_asymp}), one obtains $\ell_{\rm max}=6$ for M0, $\ell_{\rm max}=4$ for M1, and $\ell_{\rm max}=2$ for M2. 

\end{document}